\documentclass[aps,showkeys,showpacs,superscriptaddress,pra,nofootinbib,10pt,notitlepage]{revtex4-1}
\usepackage{graphicx}  % needed for figures
\usepackage{amsmath,amsfonts,bbm,MnSymbol}   % for math
\usepackage{amsfonts}
\usepackage{enumitem}	% allows removal of spaces between items in enumerate environment
\usepackage[breaklinks=true,colorlinks=true,linkcolor=blue,urlcolor=blue,citecolor=blue]{hyperref}
\usepackage{comment}
\usepackage{color}
\usepackage[makeroom]{cancel}
\pdfoutput=1

\usepackage{dsfont} % allows mathds

\renewcommand{\[}{\begin{equation}}
\renewcommand{\]}{\end{equation}}

\begin{document}

\title{Relativistic quantum clocks}	

\author{Maximilian P. E. Lock}
\email{maximilian.lock12@imperial.ac.uk}
\address{Department of Physics, Imperial College, SW7 2AZ London, United Kingdom}
\address{Faculty of Physics, University of Vienna, Boltzmanngasse 5, 1090 Vienna, Austria}
\author{Ivette Fuentes}\thanks{Previously known as Fuentes-Guridi and Fuentes-Schuller.}
\affiliation{Faculty of Physics, University of Vienna, Boltzmanngasse 5, 1090 Vienna, Austria}
\affiliation{School of Mathematical Sciences, University of Nottingham, University Park,
Nottingham NG7 2RD, United Kingdom}

\date{\today}

\begin{abstract}
The conflict between quantum theory and the theory of relativity is exemplified in their treatment of time. We examine the ways in which their conceptions differ, and describe a semiclassical clock model combining elements of both theories. The results obtained with this clock model in flat spacetime are reviewed, and the problem of generalizing the model to curved spacetime is discussed, before briefly describing an experimental setup which could be used to test of the model. Taking an operationalist view, where time is that which is measured by a clock, we discuss the conclusions that can be drawn from these results, and what clues they contain for a full quantum relativistic theory of time.
\end{abstract}

\maketitle

\tableofcontents

\section{Time in quantum mechanics and general relativity} \label{sTimeInQMGR}

When an experiment is carried out, the experimenter hopes to gain some information about nature through her controlled interaction with the system under study. In classical physics, systems posess a set of measurable properties with definite values, which can in principle be interrogated simultaneously, to arbitrary accuracy, and without affecting the values of those properties. Any uncertainty in the measurements arises from some lack of knowledge on the part of the experimenter (for example due to some imperfect calibration of the aparatus) which could, in principle, be corrected. In quantum theory on the other hand, uncertainty relations between conjugate variables, and the necessary backreaction of the measurement on the system, combine to pose strict limits on the information which can be obtained from nature. Although there are nontrivial complications in defining time as a quantum observable (see the introductory discussion in~\cite{giannitrapani1997positive}, and Section~12.8 of~\cite{peres2006quantum}, for example), it is nonetheless apparent that quantum restrictions must also be applied to its measurement~\cite{mandelstam1945uncertainty,peres1980measurement,braunstein1996generalized}.

The general theory of relativity (GR) lies within the classical paradigm with respect to the measurements that can be performed, though the outcomes of such measurements are affected by the observer's state of motion, and the distribution of energy around them. The theory is built upon the notion of ``ideal'' clocks and rods, through which the observer gathers information. In special relativity, an ideal clock is a pointlike object whose rate with respect to some observer depends only on its instantaneous speed, and not directly on acceleration~\cite{rindler2006relativity}. The latter property is sometimes referred to as the ``clock postulate'', and can be justified by the fact that an observer can ``feel'' their own acceleration, in contrast to velocity. Therefore, given a clock whose rate depends on acceleration in a well-defined manner, one can simply attach an accelerometer to it, and use the resulting measurements to add/subtract time such that the acceleration effect is removed, recovering an ideal clock. Combining this clock postulate with the constancy of the speed of light, one finds that an ideal clock measures the proper time along its trajectory according to the usual formulas of special relativity. The concept of an ideal clock (and therefore proper time) is imported into GR via Einstein's equivalence principle~\cite{rindler2006relativity}. This principle states that local experiments conducted by a freely-falling observer cannot detect the presence or absence of a gravitational field. Here ``local'' means within a small enough volume that the gravitational field can be considered uniform.

We note four conceptual issues which arise when combining GR and quantum theory. The first is understanding how quantum theory imposes constraints on the clocks and rods of GR, and how this in turn affects the information gathered by an observer. Here, we concern ourselves with clocks, and we refer the reader to~\cite{hossenfelder2013minimal} for a review of possible limitations to spatial measurements. Some progress has been made with this issue, for example~\cite{salecker1958quantum}, wherein the mass and mass uncertainty of a clock system are related to its accuracy and precision (neglecting spacetime curvature). In~\cite{burderi2016quantum}, using a gedankenexperiment, one such mass-time relation is rederived and combined with the ``hoop conjecture'' (a supposed minimum size before gravitational collapse~\cite{misner1973gravitation}), to argue that the product of a clock's spatial and temporal uncertainty is bounded below by the product of the Planck length and the Planck time.

A second, perhaps more difficult problem, is that of reconciling the definition of time via a pointlike trajectory in GR with the impossiblity of such trajectories according to quantum mechanics (a result of the uncertainty principle between position and momentum). A third issue is the prediction that acceleration affects quantum states via the Unruh~\cite{fulling1973nonuniqueness,unruh1976notes} and dynamical Casimir~\cite{moore1970quantum} effects (DCE), which in turn will affect clock rates~\cite{lorek2015ideal}. One must therefore reconsider whether it is always possible to measure and remove acceleration effects and recover an ideal clock. Finally, the fourth issue is that, given the locality of the equivalence principle (i.e. it only holds exactly when we consider a pointlike observer), it is unclear to what extent it applies to quantum objects, which do not follow pointlike trajectories.

We investigate the interplay of these four issues, seeking to answer the following questions: what time does a quantum clock measure as it travels through spacetime, and what factors affect its precision? What are the fundamental limitations imposed by quantum theory on the measurement of time, and are these affected by the motion of the clock? To answer this question, we cannot in general rely on the Schr\"{o}dinger equation, as we must use a particular time parameter therein, which in turn requires the use of a particular classical trajectory.\footnote{Ignoring this difficulty, and naively picking some time coordinate, one finds that the Schr\"{o}dinger equation for a free particle does not possess the necessary symmetry; it is invariant under Galilean (rather than Lorentz) transformations.} The relativistic clock model detailed in Section~\ref{sRelQuanClock} gives a compromise; its boundaries follow classical trajectories, but the quantum field contained therein, and hence the particles of that field, do not. In Section~\ref{sConc} we examine the extent to which this clock has allowed the four issues discussed above to be addressed, and possible future progress.

Given the difference in the scales at which quantum theory and GR are usually applied, one may ask what we expect to gain by examining their overlap. Our response to such a question is threefold. Firstly, we note that optical clocks have reached a precision where gravitational time dilation as predicted by GR has been measured over scales accessible within a single laboratory~\cite{chou2010optical}. Indeed, modern clocks are precise enough that they are sensitive to a height change of 2~cm at the Earth's surface~\cite{nicholson2015new}. Given the rate of improvement of this technology (see Figure~1 of~\cite{poli2013optical}, for example), one can anticipate an even greater sensitivity in the near future. The detection of a nuclear transition in thorium-229~\cite{von2016direct}, proposed as a new frequency standard~\cite{campbell2012single}, means that we may soon enter an era of ``nuclear clocks'', surpassing that which is achievable with clocks based on electronic transitions. Considering this ever-increasing precision together with proposals to exploit quantum effects for superior timekeepking (e.g.~\cite{komar2014quantum,hosten2016measurement}), we argue that a consideration of GR alongside quantum theory will become not simply possible, but in fact necessary in order to accurately describe the outcomes of experiments.

Our second response is to point out the possibility of new technologies and experiments. There are already suggestions exploiting the clock sensitivity mentioned above, such as the proposal to use changes in time dilation for earthquake prediction and volcanology~\cite{bondarescu2015potential}. On the other hand, there are proposals to use effects which are both quantum and relativistic in order to measure the Schwarzschild radius of the Earth~\cite{bruschi2014quantum}, or to make an accelerometer~\cite{dragan2011quantum}, for example. See~\cite{howl2016gravity} for a review of experiments carried out or proposed which employ both quantum and general relativistic features. Beyond specific proposals, there are practical questions which we cannot answer with quantum mechanics and GR separately; for example, what happens if we distribute entanglement across regions with differing spacetime curvatures, or how do we correlate a collection of satellite-based quantum clocks? The answers to these questions are relevant for proposals to use correlated networks of orbiting atomic clocks for entanglement-assisted GPS~\cite{komar2014quantum}, or to search for dark matter~\cite{derevianko2014hunting}

Finally, there is a strong motivation from the perspective of fundamental science to investigate the nature of time at the overlap of GR and quantum theory. Beyond the intrinsic interest of finding a coherent combination of the two most fundamental theories in physics, a quantum relativistic conception of time may be of relevance when using quantum clocks to test the equivalence principle~\cite{gabriel1990testing,davies2004quantum} and to single out GR from the family of gravitational theories obeying this principle~\cite{reynaud2009testing}, for example. In addition, since we expect a viable theory of quantum gravity to also be a quantum theory of space and time, it must either reproduce a relativistic quantum theory of time in the semiclassical limit, or contradict it, giving a potential test of the quantum gravity theory compared to the semiclassical one that we use here.

\section{A semiclassical approach: quantum field theory in curved spacetime}

To answer the questions raised in Section~\ref{sTimeInQMGR}, a framework incorporating elements of both quantum mechanics and general relativity is needed. At high energies, one would need a full theory of quantum gravity to do this. However, if we only consider the energy scales accessible in current (or near-future) experiments, and where the spacetime curvature is relatively low, we can employ the semiclassical methods of quantum field theory in curved spacetime (QFTCS). It is semiclassical in the sense that quantum matter and radiation are embedded in a classical curved spacetime, the latter being subject to Einstein's equations. QFTCS also allows us to describe quantum fields from the perspective of non-inertial observers, leading to predictions of novel phenomena related to acceleration, namely the Unruh effect and the DCE, mentioned in Section~\ref{sTimeInQMGR}. The latter effect has been demonstrated experimentally~\cite{wilson2011observation,lahteenmaki2013dynamical}, as we briefly describe in Section~\ref{sImplementations}. It is worth underlining that these effects are both quantum mechanical and relativistic in nature, and cannot be derived by, for example, simply inserting a relativistic proper time into the Schr\"{o}dinger equation of quantum mechanics. To fully include (classical) relativity into the quantum dynamics, one needs QFTCS.

In recent years, aspects of quantum information have been incorporated into QFTCS in a collection of research efforts known as relativistic quantum information. This has allowed, for example, investigations into the effect of spacetime dynamics~\cite{ball2006entanglement,fuentes2010entanglement} and non-inertial motion~\cite{fuentes2005alice,alsing2012observer,friis2012motion,bruschi2012voyage} on quantum entanglement, and the potentially detrimental~\cite{adesso2007continuous,friis2013relativistic} or advantageous~\cite{friis2012quantum,bruschi2013relativistic} consequences of such motion for some quantum information applications.

A particularly fruitful branch of relativistic quantum information is the incorporation of quantum metrology into a relativistic setting~\cite{ahmadi2014quantum,ahmadi2014relativistic}, with a number of possible applications including the measurement of the Schwarzschild radius of the Earth~\cite{bruschi2014quantum} and the detection of gravitational waves in small-scale BEC experiments~\cite{sabin2014phonon}. The application of relativistic quantum metrology to the measurement of time is the subject of Section~\ref{sRelQuanClock}.

\section{A relativistic quantum clock} \label{sRelQuanClock}

\subsection{The clock model} \label{sClockModel}
The clock model introduced in~\cite{lindkvist2014twin} allows us to integrate aspects of both general relativity and quantum mechanics. It consists of a particular mode of a localized quantum field; the boundaries confining the field define the spatial extent of the clock, and the clock time is given by the phase of a single-mode Gaussian state. This gives a clock that can undergo classical relativistic trajectories, but whose dynamics are described by QFTCS. The former property means that we can compare this to a pointlike clock by considering a classical observer following the trajectory of the center of the cavity, while the latter property allows us to consider the effect of the spacetime curvature on the whole extent of the quantum field, instead of relying on the Schr\"{o}dinger equation. The transformation of the quantum state of a localized field due to boundary motion is a well-studied problem in flat spacetime~\cite{bruschi2012voyage,bruschi2013mode}, particularly the generation of particles due to the DCE~\cite{moore1970quantum}. Since the frequencies of the field modes depend on the length between the boundaries, one must be careful to choose the trajectories in such a way that the comparison with the pointlike classical clock is a fair one. One must also be careful to distinguish between classical effects arising purely from the spatial extent of the clock, and novel quantum effects due to mode-mixing and particle creation. % Note that the best modern clock systems, atomic clocks, ultimately rely upon an estimation of the phase of an electromagnetic field state? Also note that this is the quantum version of the Einstein's light clock

To analyze the effect of non-inertial motion and spacetime curvature on the clock, we first need to describe their effect on its quantum state, giving us the change in phase (i.e. clock time). Since the phase is subject to an quantum uncertainty relation with respect to the particle number (see~\cite{opatrny1995number}, for example), a change in the state of the field will in general modify the precision with which the phase can be estimated. Once these changes have been determined, one can compare the overall phase with the corresponding classical result to find quantum relativistic shifts in the clock time, and one can see how the precision of the clock is affected by considering the change in phase estimation precision. Before discussing the results obtained using this clock model, we give a brief overview of the framework underpinning it.

\subsection{Theoretical framework} \label{sFramework}

\subsubsection{A localized quantum field in curved spacetime} \label{sQFTCS}
The simplest quantum field theory is that of the massless scalar field. This can be used, for example, to approximate the electromagnetic field when polarization can be ignored~\cite{friis2013scalar}, or phononic excitations in a proposed relativistic BEC setup~\cite{fagnocchi2010relativistic}. For simplicity, we consider one spatial and one temporal dimension. In a general $1+1$D spacetime, the massless scalar field satisfies the Klein-Gordon equation~\cite{birrell1984quantum}
\begin{equation} \label{eKG}
\Box \Phi = 0, \qquad \text{ with } \qquad \Box := g^{\mu \nu} \nabla_{\mu} \nabla_{\nu}
\end{equation}
In some coordinate system $(t,x)$, imposing the boundary conditions $\Phi(t,x_{1})=0$ and $\Phi(t,x_{2})=0$ for a given $x_{1}$ and $x_{2}$, we describe either an electromagnetic field in cavity or the phonons of a BEC trapped in an infinite square well. After finding a set of mode solutions to Equation~\ref{eKG}, which we denote $\phi_{m}(t,x)$, one can (under certain conditions, discussed briefly in Section~\ref{sCurvedResults}) associate particles with the modes, and quantize the field by introducing creation and annihilation operators $a_{m}^\dag$ and $a_{m}$. These satisfy the usual bosonic commutation relations, $\left[ a_{m}^\dag , a_{n} \right]=\delta_{mn}$, and can be used to define the vacuum and Fock states in the usual way. The total scalar field is then given by
\begin{equation} \label{eTotalField}
\Phi(t,x) = \sum_{m} \left[ a_{m} \phi_{m}(t,x) +  a_{m}^\dag \phi_{m}^{*}(t,x)  \right] .
\end{equation}
If the field can be described in terms of a second set of mode solutions, we can relate these to the first set by means of a Bogoliubov transformation. Denoting the creation and annihilation operators associated with the new set of solutions by $b_{m}$ and $b_{m}^\dag$, the Bogoliubov transformation can be written as
\begin{equation} \label{eBogoTrans}
b_{m} = \sum_{n} \left( \alpha_{mn}^{*} a_{n} - \beta_{mn}^{*} a_{n}^\dag \right) ,
\end{equation}
where $\alpha_{mn}$ and $\beta_{mn}$ are known as the Bogoliubov coefficients, and can be computed using an inner product between the first and second set of mode solutions (see~\cite{birrell1984quantum} for details). These transformations can be used, for example, to represent changes in coordinate system between inertial and non-inertial observers, or the effect of Gaussian operations or of spacetime dynamics. Mixing between modes due to the transformation is determined by the $\alpha_{mn}$, while the $\beta_{mn}$ correspond to the generation of particles. The fact that the $\beta_{mn}$ are non-zero for Bogoliubov transformations between inertial and non-inertial observers leads to the Unruh effect and the DCE.

\subsubsection{The covariance matrix formalism} \label{sCovMatForm}
The relativistic clock model described in Section~\ref{sClockModel} makes use of only a single mode of the field after the transformation. It is then very advantageous to work with the covariance matrix formalism, which greatly simplifies the process of taking a partial trace over field modes. In doing so, we restrict ourselves to the consideration of Gaussian states of the field. The set of such states is closed under Bogoliubov transformations. Defining the quadrature operators for mode $n$ by $X_{2n-1}:=\frac{1}{2}\left( a_{n}+a_{n}^\dag \right)$ and $X_{2n}:=-\frac{i}{2}\left( a_{n}-a_{n}^\dag \right)$, a Gaussian state is completely determined by the first moments $q^{(n)}:=\langle X_{2n-1} \rangle$ and $p^{(n)}:=\langle X_{2n} \rangle$, and the second moments i.e. the covariance matrix
\begin{equation}
\sigma_{ij} = \frac{1}{2} \left\langle \left\lbrace X_{i} , X_{j} \right\rbrace \right\rangle - \left\langle X_{i} \right\rangle \left\langle X_{j} \right\rangle .
\end{equation}
To take a partial trace over some modes, one simply removes the corresponding rows and columns from the covariance matrix. Let $k$ and $\sigma^{(k)}$ denote respectively a mode of interest and the reduced covariance matrix of that mode. Now consider some initial state with first moments $q^{(k)}_{0}$ and $p^{(k)}_{0}$, and reduced covariance matrix $\sigma^{(k)}_{0}$. After a Bogoliubov transformation, the first and second moments are given by~\cite{ahmadi2014relativistic,ahmadi2014quantum}
\begin{equation} \label{eCMBogoMoms}
       \begin{pmatrix} q^{(k)} \\ p^{(k)} \end{pmatrix} = \mathcal{M}_{kk} \begin{pmatrix} q^{(k)}_{0} \\ p^{(k)}_{0} \end{pmatrix} \qquad \text{ and } \qquad 
       \sigma^{(k)} = \mathcal{M}_{kk} \sigma^{(k)}_{0} \mathcal{M}_{kk}^{T} + \frac{1}{4} \sum_{n \neq k} \mathcal{M}_{kn} \mathcal{M}_{kn}^{T},
\end{equation}
with
\begin{equation} \label{eCMBogoMat}
\mathcal{M}_{mn} :=
\begin{pmatrix}
	\Re \left( \alpha_{mn} - \beta_{mn} \right)  & \Im \left( \alpha_{mn} + \beta_{mn} \right) , \\
	-\Im \left( \alpha_{mn} - \beta_{mn} \right) & \Re \left( \alpha_{mn} + \beta_{mn} \right) 
\end{pmatrix}
\end{equation}

A single-mode Gaussian state is also characterized by the following parameters: the (real) displacement $\alpha$, the (complex) squeezing $\xi=r e^{i \phi}$, the phase $\theta$ and the purity $P$. These parameters can be expressed in terms of the first and second moments as follows
\begin{subequations} \label{eGaussStateParams}
\begin{align}
       & \alpha 		= \sqrt{q^{(k)2}+p^{(k)2}}	,\\[1.6mm]
       & \tan \theta	= \frac{p_{(k)}}{q_{(k)}}, \\[1mm]
       & P			= \frac{1}{4 \sqrt{\mathrm{det} \, \sigma_{(k)}}}, \\
       & r	= \frac{1}{2} \mathrm{arctanh} \left( \frac{\sqrt{\left(\sigma^{(k)}_{11}-\sigma^{(k)}_{22}\right)^{2}+\left( 2\sigma^{(k)}_{12} \right)^{2}}}{\sigma^{(k)}_{11}+\sigma^{(k)}_{22}} \right), \\
	   & \tan (2\theta+\phi) 	= \frac{2 \sigma^{(k)}_{12}}{\sigma^{(k)}_{11}-\sigma^{(k)}_{22		}}.
\end{align}
\end{subequations}

\subsubsection{Relativistic quantum metrology}
The field of quantum metrology developed in parallel to quantum information~\cite{giovannetti2011advances,wiseman2009quantum}, and is concerned with the application of quantum features, such as squeezing or entanglement, to improve the precision with which some quantity is measured. Say we seek to estimate a parameter $\lambda$ by making $M$ measurements. The variance $\Delta \lambda$ of estimators of $\lambda$ satisfy the quantum Cram\'{e}r-Rao bound~\cite{giovannetti2011advances}
\begin{equation}
\Delta \lambda \geq \frac{1}{\sqrt{M H_{\lambda}}},
\end{equation}
where $H_{\lambda}$ is the quantum Fisher information (QFI). We can therefore use the QFI to quantify the precision with which a parameter can be measured: a greater QFI implies a greater precision. We note, however, that the QFI is obtained by an unconstrained optimization over all generalized measurements~\cite{giovannetti2011advances}, and as such gives the theoretical maximum precision, without any consideration of the feasibility of the measurement process required to achieve it.

In recent years there has been an interest in using squeezed light to improve the sensitivity of gravitational measurements such as in the LIGO gravitational wave detector~\cite{aasi2013enhanced}, and in atom interferometric measurements of gravitational field gradients~\cite{szigeti2014squeezed}. Typically, proposals consider non-relativistic quantum theory and Newtonian physics, while others include some corrections due to GR~\cite{altschul2015quantum}. In~\cite{ahmadi2014relativistic,ahmadi2014quantum}, quantum metrology was considered using QFTCS, giving a more fully relativistic application of quantum metrology. Applying these ideas, we consider $\lambda$ to be encoded into the Bogoliubov coefficents, and thus into the matrices $\mathcal{M}_{mn}$ given by Equation~\ref{eCMBogoMat}. From the corresponding transformation of the first and second moments (Equations~\ref{eCMBogoMoms}), and the expression of the Gaussian state parameters in terms of these moments (Equations~\ref{eGaussStateParams}), one can see how the parameters encode $\lambda$. We apply quantum metrology to the estimation of the phase of a single-mode Gaussian state, i.e. $\lambda=\theta$. The QFI for the phase, written in terms of the other Gaussian state parameters, is given by~\cite{lindkvist2015motion}
\begin{equation} \label{eQFIphase}
H_{\theta} = 4 \alpha^{2} P \left[ \cosh (2r) + \sinh (2r) \cos \phi \right] + \frac{4 \sinh^{2} (2r)}{1+P^2}.
\end{equation}

\subsection{The effect of non-inertial motion} \label{sFlatResults}
To describe an accelerating clock, one can make use of so-called Rindler coordinates. These coordinates are natural for describing accelerated observers in a number of ways. For example, an observer at any fixed spatial Rindler coordinate experiences undergoes a constant proper acceleration and has a proper time linearly proportional to the Rindler time coordinate. Furthermore, an extended object which is stationary in Rindler coordinates satisfies a number of desirable properties, including Born rigidity~\cite{born1909theory} and a constant ``radar length'' (the length as measured by the round-trip-time of a light pulse)~\cite{rindler2006relativity}. By judiciously connecting together Rindler coordinates corresponding to different proper accelerations, the Bogoliubov transformation corresponding to a continuously varying (finite-duration) proper acceleration can be calculated~\cite{bruschi2013mode}. For the results described in this section however, it suffices to join segments of constant proper acceleration with segments of inertial motion, as detailed in~\cite{bruschi2012voyage}. 

\begin{figure}[h]
  \centering
  \includegraphics[width=0.5\textwidth]{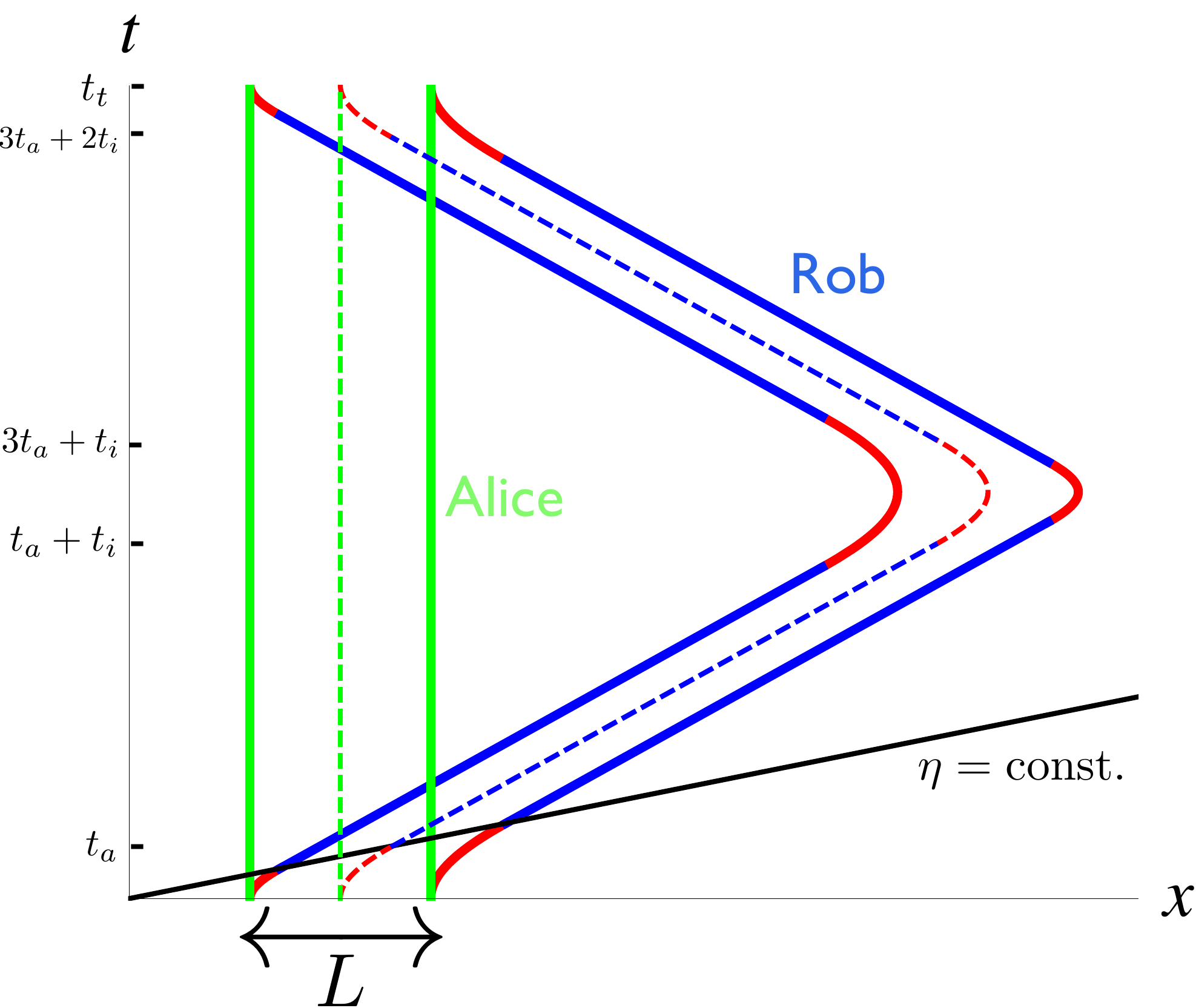}
  \caption{The twin paradox trajectory. One ``twin'', Alice, remains stationary in some inertial reference frame while the other, Rob, undergoes a round trip. Rob's trajectory consists of segments of proper acceleration of magnitude $a$ (red) and segments of inertial motion (blue). The dashed lines give the trajectories of the corresponding pointlike observers. Figure taken from~\cite{lindkvist2014twin}.} \label{fTwinParaTraj}
\end{figure}
\begin{figure}[h]
  \centering
  \includegraphics[width=0.5\textwidth]{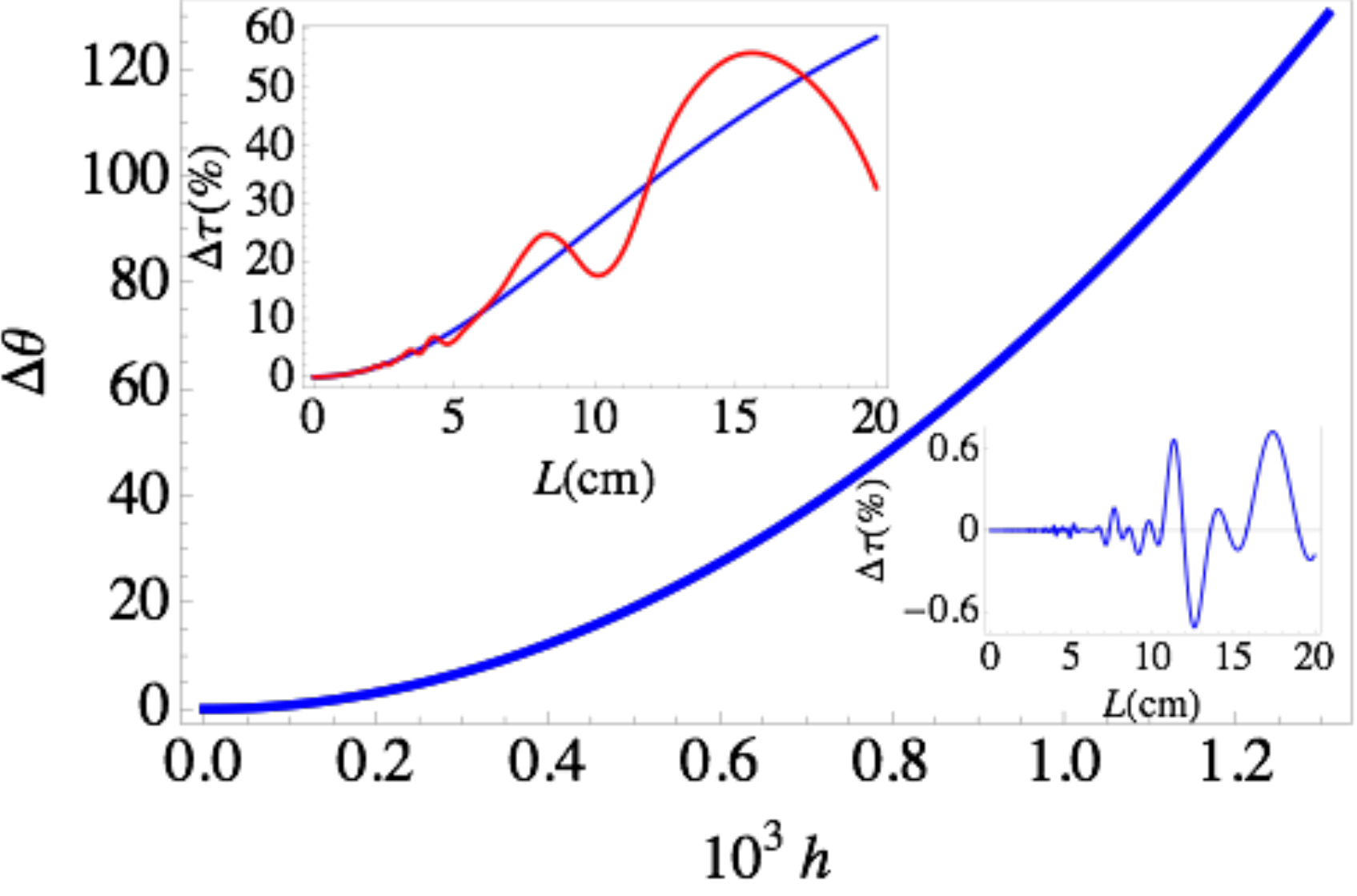}
  \caption{Time dilation, classical and quantum relativistic acceleration effects using feasible parameters for a SQUID setup, repeating the scenario 500 times. Unless variable, the parameters are $t_{a}=1$~ns, $t_{i}=0$~ns, $L=1.1$~cm, and $a=1.7 \times 10^{15}$~m/s${}^{2}$. \textbf{Main plot:} phase difference between the twins, using spatially-extended relativistic quantum clocks ($h \coloneq a L/c^{2}$). \textbf{Left inset:} time difference between Rob using a pointlike and a spatially extended clock, with (red) and without (blue) mode-mixing and particle-creation effects, as a percentage of the total time dilation between the twins. \textbf{Right inset:} percentage of the total time dilation between twins due exclusively to particle-creation. Figure taken from~\cite{lindkvist2014twin}.} \label{fTwinParaDev}
\end{figure}

In~\cite{lindkvist2014twin}, the effect of non-inertial motion on the clock time was investigated in the famous twin-paradox scenario. In this scenario, one clock remains motionless while another undergoes a round trip, and the stationary clock registers more time passing than the round-trip clock. The round-trip trajectory was composed of periods of constant proper acceleration $a$ interspersed with periods of inertial motion (see Figure~\ref{fTwinParaTraj}, and reference~\cite{lindkvist2014twin} for more details). The clocks were initialized in the same coherent state. First considering the purely classical deviation (i.e. in the absence of mode-mixing and particle creation), between a pointlike and a spatially extended clock, one finds a difference only during the periods of acceleration. During a period of proper acceleration, the time measured by the cavity-clock $\tau_{cav}^{a}$ can be related to the proper time of a pointlike observer $\tau_{point}^{a}$ by
\begin{equation} \label{eClassLengthEffect}
\frac{\tau_{cav}^{a}}{\tau_{point}^{a}} = 1 - \frac{1}{12} \left( \frac{a L}{c^{2}} \right)^{2} + \mathcal{O} \left( \left( \frac{a L}{c^{2}}  \right)^{4} \right)
\end{equation}
Recalling that less time passes for the accelerated pointlike ``twin'' than the stationary one, we see from Equation~\ref{eClassLengthEffect} that the classical effect of the clock's nonzero spatial extent is to increase this disparity. If we now include mode-mixing and particle creation effects due to the motion, as determined by the Bogoliubov transformation, we find a non-trivial relation between the time as measured by the relativistic quantum clock model and a pointlike clock. This is illustrated in Figure~\ref{fTwinParaDev} using experimentally feasible parameters for the superconducting quantum interference device (SQUID) setup discussed in Section~\ref{sImplementations}. The left inset of Figure~\ref{fTwinParaDev} shows the difference between the quantum clock and a pointlike clock, both with and without mode-mixing and particle-creation effects, as a function of the clock size $L$. The right inset gives the percentage of the effect due to particle creation alone, again as a function of the clock size. Particle creation being a purely quantum effect, this gives a new quantum contribution to the relativistic phenomenon of time dilation. The complicated oscillatory behavior of this contribution is due to the non-trivial $L$-dependence of numerous complex terms which are added together to give the relevant Bogoliubov coefficients (see the appendix of~\cite{lindkvist2014twin} for details). The main plot of Figure~\ref{fTwinParaDev} gives the relative phase shift between the twins' quantum clocks.

In~\cite{lindkvist2015motion}, the effect of non-inertial motion on the precision of the clock was investigated. This depends on the state in which the clock is initialized. The QFI for the phase of a Gaussian state was given in Equation~\ref{eQFIphase}. From this we see that, for $\phi \notin (\pi / 2, 2 \pi/3)$ and a given purity, the precision of phase estimation increases with the real displacement parameter $\alpha$ and the magnitude $r$ of the squeezing. For a given average particle number $\langle N \rangle$, the squeezed vacuum state is the best Gaussian state for phase estimation~\cite{monras2006optimal}. 
\begin{figure}[h]
  \centering
  \includegraphics[width=0.8\textwidth]{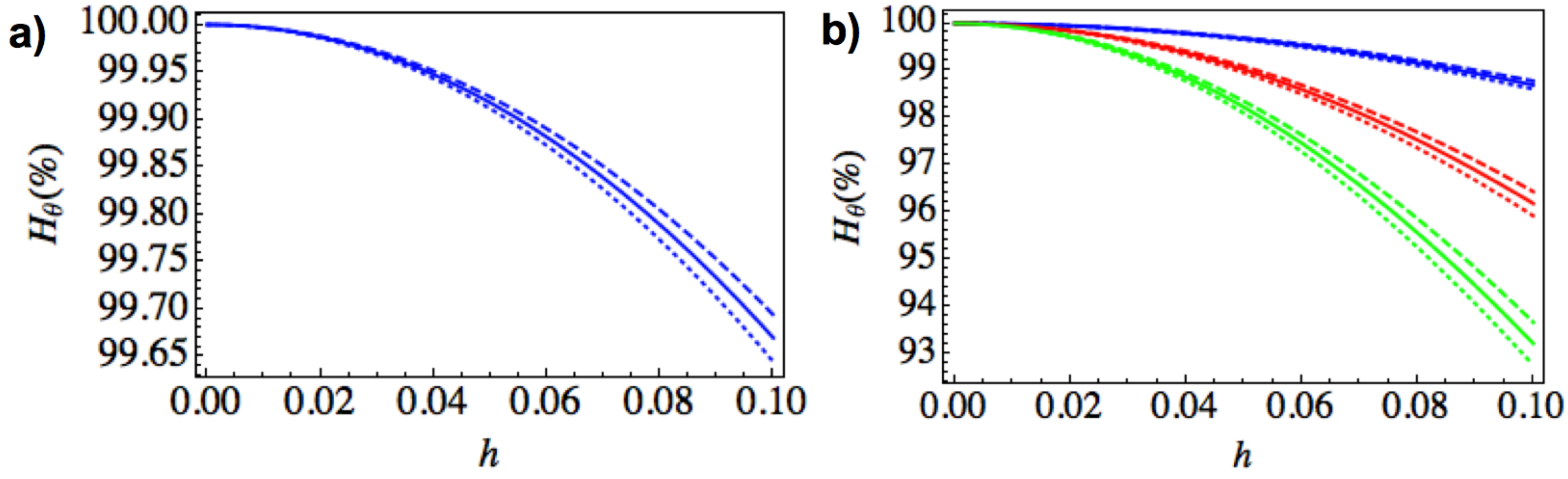}
  \caption{The change in the QFI (given as a percentage of its pre-motion value) after non-inertial motion with $h \coloneq a L/c^{2}$, for a) a coherent initial state, and b) a squeezed vacuum initial state with $\langle N \rangle=1$ (blue), $\langle N \rangle=5$ (red) and $\langle N \rangle=10$ (green). The phase accrued during each $t_{a}$ of acceleration was $\theta_{a}=\pi$. The solid curves give the effect of mode-mixing alone, while the dotted and dashed curves incorporate the effect of particle creation for an initial phase of $\theta_{0}=0$ and $\theta_{0}=\pi/2$ respectively. Figure taken from~\cite{lindkvist2015motion}.} \label{fMotPrec}
\end{figure}
In Figure~\ref{fMotPrec}, the effect of non-inertial motion on the QFI for coherent and squeezed vacuum states is depicted. In particular, one can see the separability of the mode-mixing and particle creation effects. Mode-mixing acts to decrease the QFI, and therefore the precision of the clock, more so for the squeezed vacuum than for the coherent state, though in the regime considered there is no point at which the coherent state gives a better clock than the squeezed vacuum. Particle creation, on the other hand, can either ameliorate or exacerbate this effect, depending on the initial phase $\theta_{0}$ of the clock. For large $\langle N \rangle$, the degradation due to mode-mixing dominates, but as $\langle N \rangle$ decreases, one arrives at a regime where particle-creation effects dominate. For low enough $\langle N \rangle$ and a careful choice of parameters one can even find cases where the QFI is improved as a result of the generation of the appropriate squeezing, though the set of such cases is relatively small. One can therefore conclude that the typical effect of non-inertial motion is to decrease the precision of the clock.

\subsection{Generalizing to curved spacetime} \label{sCurvedResults}

For a pointlike observer, Einstein's equivalence principle allows us to equate free-fall with flat spacetime. However, for a system with some finite extent in a gravitational field, tidal forces will reveal the curvature of the spacetime. Likewise, one can equate a pointlike object at rest in a gravitational field with one undergoing some proper acceleration in flat spacetime, and one finds again that this equivalence breaks down for a system with finite extent. This is illustrated in~\cite{desloge1989nonequivalence}, for example, where it is shown that a reference frame at rest in a uniform gravitational field is not equivalent to a uniformly accelerating one. Given these considerations, when seeking to apply the results discussed above to curved spacetimes, one can only invoke the equivalence principle in a limited sense. Here, we illustrate this in the Schwarzschild spacetime, though a similar argument can be applied to any static spacetime.

In the work discussed in previous sections, Rindler coordinates were used to represent the accelerated observer. One example of such coordinates, $(\eta,\chi)$, can be obtained from inertial coordinates $(T,X)$ by the transformation
\begin{equation}
T = \chi \sinh \eta, \qquad X = \chi \cosh \eta .
\end{equation}
Considering a set of observers fixed at each spatial Rindler coordinate $\chi$, we obtain a particular profile of constant proper accelerations experienced by these observers: $a_{R}=1/\chi$. Now consider the Schwarzschild spacetime corresponding to a mass $M$, in the usual Schwarzschild coordinates $(t,r)$. The metric is given by
\begin{equation}
ds^{2} = -f(r) dt^{2} + \frac{1}{f(r)} dr^{2} \qquad \text{ with } \qquad f(r) \coloneq 1-\frac{r_s}{r} \; \text{ and } \; r_{s} \coloneq \frac{2 G M}{c^{2}} .
\end{equation}
In this case, observers at fixed $r$ experience the constant proper acceleration~\cite{dahia2011static}
\begin{equation}
a_{S}=\frac{r_{s}}{2 r^{2}}\frac{1}{\sqrt{f(r)}},
\end{equation}
which is evidently different from the Rindler case. Since the clock has non-negligible extent, we cannot equate these two circumstances in general. Close to the event horizon at $r=r_{s}$ however, one can approximate the spacetime experienced by stationary Schwarschild observers using Rindler coordinates~\cite{dahia2011static}, giving an approximate equality between $a_{R}$ and $a_{S}$, and in this case one can import the method discussed in Section~\ref{sFlatResults} into an investigation in curved spacetime.

To examine more general situations, we need to be able to describe the effect of general boundary motion through curved spacetime on the quantum state of the field, a problem whose solution was unknown until recently. We gave such a solution in~\cite{lock2016dynamical}, providing a method for describing the effect of a finite period of cavity motion through a static curved spacetime for a broad class of trajectories. This provides us with the means to explore the effect of gravity on the clock, namely how deviations from the proper-time prescription of relativity depend on the spacetime curvature, and how the precision of the clock is affected.

There remain, however, certain challenges. In the flat spacetime case, there was an unambiguous notion of length which could be adopted, determined by demanding that an observer accelerating with the clock measure a constant length. This results in a number of desirable properties, such as Born rigidity (a lack of stresses on the clock support system), constant radar distance (the distance as measured by timing classical light pulses), and constant proper length. In curved spacetime, however, such notions do not necessarily coincide, and there is no unambiguous generalization of Rindler coordinates. Fermi-Walker coordinates are a candidate for such a generalization, but it unclear if this theoretical construction is in keeping with the operationalism which we have until now adopted (for example by defining time as that which is measured by a clock). We are currently investigating different notions of length in curved spacetime, and how the choice of which notion to adopt affects the measurement of time.

The discussion above considered only static spacetimes. Now including the possibility of non-static ones, we can ask how the spacetime dynamics themselves affect the clock. This question brings with it an added complication: in order to associate a set of solutions to the field equations with particle modes, we require that the spacetime admits a timelike Killing vector field, which is by no means guaranteed for a nonstationary spacetime. Without such a vector field, there is an ambiguity in the concept of particles~\cite{birrell1984quantum}. Nonetheless, there are some cases in which these issues can be overcome, such as in the usual calculation of particle creation due to an expanding universe~\cite{parker1966creation,parker2012particle}, leaving us free to apply the quantum clock model.

\subsection{Physical implementation} \label{sImplementations}

As noted in Section~\ref{sQFTCS}, the scalar field used in the clock model described above can represent light in an optical cavity (neglecting polarization), or the phonons of a BEC under certain conditions~\cite{fagnocchi2010relativistic}. We only consider the former implementation here. Subjecting the mirrors of an optical cavity to the necessary non-inertial motion\footnote{Note that it is not acceleration but rather its time-derivative (the ``jerk'') which produces the effect~\cite{fulling1976radiation,ford1982quantum}.} is technically infeasible~\cite{braggio2005novel}. 
\begin{figure}[h]
  \centering
  \includegraphics[width=0.55\textwidth]{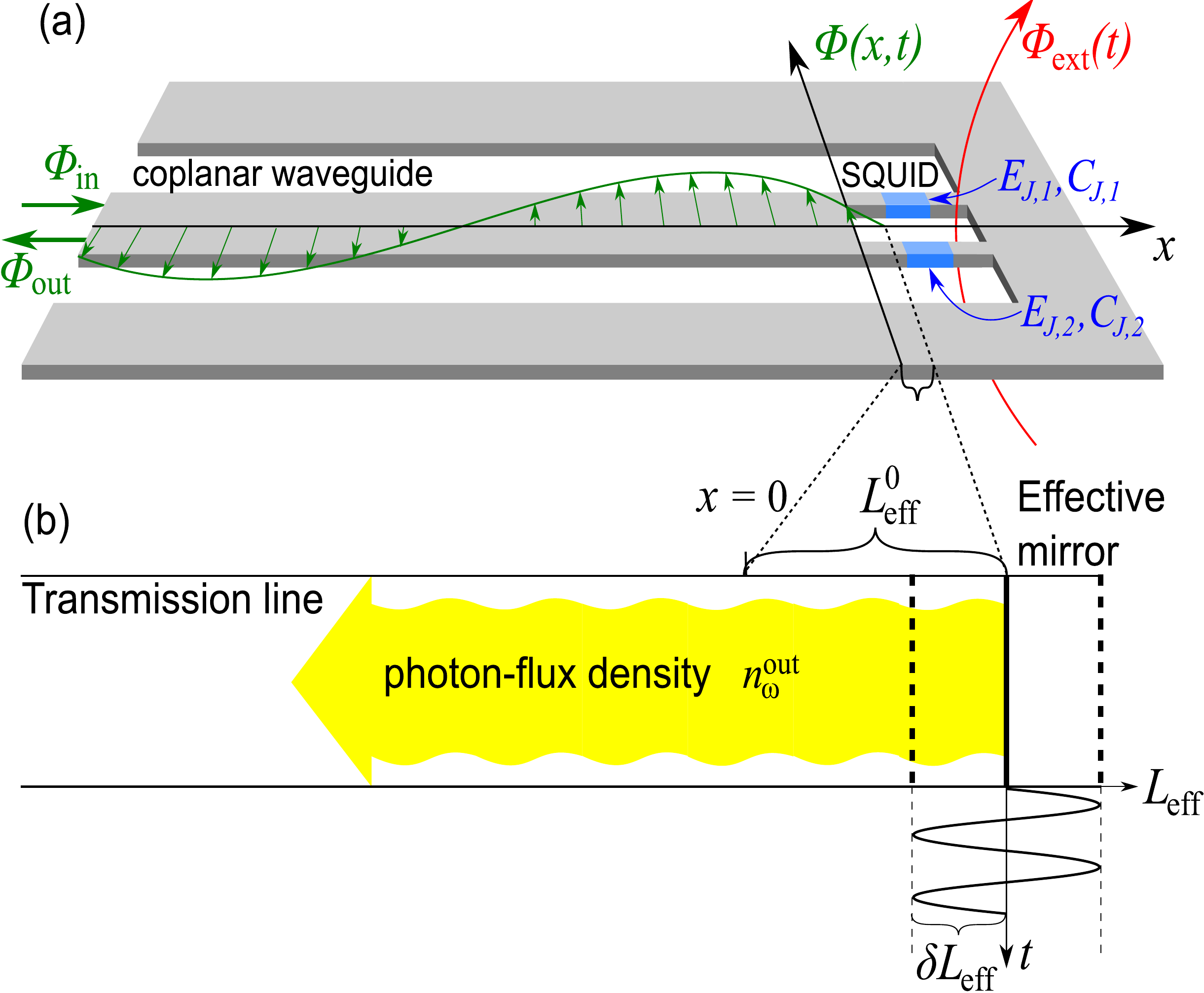}
  \caption{a) Depiction of a waveguide terminated by a SQUID, with the external magnetic flux $\Phi_\text{ext}$ determining the boundary condition. b) The corresponding transmission line and effective mirror. Figure taken from~\citep{johansson2009dynamical}.} \label{fSQUIDsetup}
\end{figure}
To circumvent this requirement, a novel solution was proposed in~\cite{johansson2009dynamical}; by placing a SQUID at one or both ends of a waveguide, one can create effective mirrors whose position is determined by the inductance of the SQUID, which is in turn controlled by an external magnetic field. Modulating the external magnetic field therefore allows the experimenter to control the position of this effective mirror. This setup is illustrated in Figure~\ref{fSQUIDsetup}. By oscillating one mirror at a particle-creation resonance, it was used to observe the DCE for the first time~\cite{wilson2011observation}. In~\cite{lindkvist2014twin}, the authors analyzed the feasibility of implementing the trajectory detailed in Section~\ref{sFlatResults} using a SQUID setup, concluding that the experiment would be challenging but possible.

\section{Conclusion} \label{sConc}

The results discussed above demonstrate both a deviation from the proper-time presciption of relativity when one considers a quantum clock with some finite extent, and a relativistic change in the quantum uncertainty associated with its measurement of time. Though these results are so far limited to flat spacetime, the main challenge to applying the model in curved spacetime, i.e. calculating the effect of motion through curved spacetime on the localized field, has now been overcome.

In Section~\ref{sTimeInQMGR}, we noted four problems arising in the overlap of quantum mechanics and relativity, which we wish to investigate. For clarity we repeat them here, before discussing each of them in turn:
\begin{enumerate}[nolistsep]
	\item finding the constraints imposed by quantum theory on clocks in GR;
	\item reconciling the proper-time prescripton of GR with the impossibility of pointlike quantum trajectories;
	\item investigating the validity of the clock hypothesis;
	\item examining the applicability of the equivalence principle to a non-pointlike quantum clock.
\end{enumerate}

To address the first problem, the quantum uncertainty of the clock measurement was quantified using the tools of quantum metrolgy, and in particular the Cramer-Rao bound. One finds that the change in precision due to relativistic motion depends upon the quantum state in which the clock was initialized, as one might expect. While some states were more robust than others, except for very particular circumstances, the motion had the effect of decreasing the QFI for all initial states, largely due to the mode-mixing. In the example considered, the more nonclassical the state, the greater its fragility with respect to the motion. A key goal of our ongoing work is to determine the effect of spacetime curvature on this analysis.

With regard to the second issue, we have attempted to move away from the proper-time prescription of GR in favor of an operationalist view, instead defining time as the result of a measurement performed on a quantum clock. This is in keeping with the Machean view that a physical theory should be based entirely on directly observable properties~\cite{barbour2001end}. We have succeeded to some extent, in that the particles of the field do not follow well defined trajectories, and the clock-time is determined by the quantum evolution of the system and not simply the length along a curve. However, we are still bound by the proper-time view, as we must choose a classical observer whose proper time parametrizes the evolution of the quantum field. Furthermore, the phase of the field, whose measurement we take as time, has a definite, noncontextual value in this model, and so is not treated as a fully quantum observable. Nonetheless, this value gives a different clock readout from the corresponding proper time, and this difference is a highly non-linear function of clock size (see the insets in Figure~\ref{fTwinParaDev}), demonstrating the non-trivial effect of the clock's non-pointlike and quantum nature.

Concerning the clock hypothesis, we can clearly state that, with the clock model employed here, one finds effects beyond the instantaneous-velocity-induced time dilation (a finding which is corroborated in~\cite{lorek2015ideal}). These effects modify both the time measured by the clock, and the precision of this measurement. This is a strong indication that, in a quantum theory of spacetime, where the latter is a measurable quantity, the clock hypothesis is not satisfied.

For the fourth problem, we discussed in Section~\ref{sCurvedResults} the applicability of the equivalence principle in the current model. To fully investigate this, we first need to study trajectories in curved spacetime. One would expect the clock to exhibit a kind of ``tidal'' effect from the difference in gravitational field across the extent of the clock system, and for this to therefore depend on the clock size and the underlying curvature. However, it seems unlikely that this will allow us to address the issue of incorporating the physical insight of the equivalence principle into a non-pointlike quantum theory.

We now note some limitations of the model and our analysis. Firstly, the QFI is obtained by optimizing over all physically allowable measurements, with no regard to their accessibility to an experimentalist, nor to the available energy. A consideration of the latter, for example its effect on the spactime which the clock measures, could result in a greater clock uncertainty.

Another potential limitation is the possibility that the results discussed here are not fundamental, but in fact particular to the specific clock model. However, the model is rather general for QFTCS: we seek a localized field, which therefore demands some kind of potential, and we justify the use of boundaries (i.e. infinite potential barriers) by noting that the shape of this potential should not play a fundamental role. One can nonetheless make this more general, by instead considering some trapping potential, or by making the boundaries only reflective to certain frequency ranges. This results in a motion-induced coupling between trapped `local' modes and global ones, the latter spanning the entire spacetime, and such a coupling would therefore likely reduce the precision of the clock. If this is true, the choice of boundaries used here can be seen as optimizing the clock precision over all possible localizing potentials.

As a final remark, we note that this clock model is, in effect, a quantum version of the common light-clock thought experiment often used to illustrate relativistic time dilation (including by Einstein himself~\cite{einstein1905elektrodynamik}).

\emph{Acknowledgements:} MPEL acknowledges support from the EPSRC via the Controlled Quantum Dynamics CDT (EP/G037043/1), and IF acknowledges support from FQXi via the `Physics of the observer' award `Quantum Observers in a Relativistic World’.

\bibliographystyle{naturemag}

\bibliography{TimeReview_BiBTeX}

\end{document}